\def\t{\textstyle}

\hsize=15cm
\vsize=22cm
\hoffset 1cm
\baselineskip 18pt
\vglue 2cm
\centerline{\bf ON THE SOLUTION AND  ELLIPTICITY PROPERTIES OF THE
SELF-DUALITY}
\centerline{\bf EQUATIONS OF CORRIGAN ET AL. IN EIGHT DIMENSIONS}
 \vskip 2cm
\centerline{Ay\c se H\"umeyra Bilge $(^*)$}
\vskip .4cm
\centerline{TUBITAK-Marmara Research Center}
\centerline{Research Institute for Basic Sciences}
\centerline{Department of Mathematics}
\centerline{P.O.Box 21, 41470 Gebze-Kocaeli, Turkey}
\centerline{e-mail:bilge@yunus.mam.tubitak.gov.tr}
\vfill
\centerline{\bf Abstract}
\baselineskip 16pt

We show that the two sets of
self-dual Yang-Mills equations in 8-dimensions
proposed in
(E.Corrigan, C.Devchand, D.B.Fairlie and J.Nuyts,
{\it Nuclear Physics} {\bf B214}, 452-464, (1983))
     form respectively elliptic and
overdetermined elliptic systems under the Coulomb gauge condition.
In the overdetermined case, the Yang-Mills fields can depend at most on $N$
arbitrary constants, where $N$ is the dimension of the gauge group.
We describe a  subvariety    ${\cal P}_8 $ of the skew-symmetric
$8\times
8$ matrices by an eigenvalue criterion and we show that the solutions of
the elliptic equations of Corrigan   et al. are among the maximal linear
submanifolds
of ${\cal P}_8$. We propose an eight order action for which the elliptic set is
a maximum.
\eject
\baselineskip 20pt

\vglue 1cm
\noindent
{\bf 1. Introduction.}

\vskip .2cm

The self-duality of a 2-form in four dimensions is defined to be the Hodge
duality. Self-dual and anti self-dual 2-forms can equivalently be
described as eigenvectors of the completely antisymmetric forth rank tensor
$\epsilon_{ijkl}$.   The latter approach is pursued by Corrigan   et al.
[Corrigan   et al., 1983)],
and self-dual 2-forms in $n$ dimensions are defined as eigenvectors of a
completely antisymmetric tensor invariant under a subgroup $G$   of $SO(n)$.
Then,  various  linear self-duality equations are obtained by specifying $G$.
In this paper we will study two sets of equations in eight dimensions arising
from invariance under
$SO(7)$. These equations denoted by {\it Set a} and {\it Set b} are given in
Section 2.

The {\it Set b} consisting of 21 equations occurs in connection with other
definitions of self-duality. The ``strongly self-dual" 2-forms
defined  in [Corrigan   et al., 1983)] are characterized by the property that their cofficients
$\omega=(\omega_{ij})$ with respect to an orthonormal basis
satisfy the equation $\omega \omega^t=\lambda I$, where $\lambda $ is a
nonzero constant.  It is shown that [Bilge   et al., 1996, Bilge, 1995], this
definition is equivalent to the self-duality definitions
of Grossman [Grossman  et al., 1984] and Trautman [Trautman, 1977] and strongly self-dual 2-forms  constitute an
$n^2-n+1$
dimensional submanifold ${\cal S}_8\cup \{0\}$ (see
Definition 3.1).  In eight dimensions the maximal
linear
submanifolds of strongly self-dual 2-forms form a six parameter family of
seven dimensional spaces, and solutions of
{\it Set b} are among these maximal linear submanifolds [Bilge  et al., 1995].

The solutions of {\it Set a} and {\it Set b} can be viewed as analogues of
self-dual 2-forms in four dimensions from different aspects.
 The strongly self-dual 2-forms, hence the solutions of {\it Set b}
saturate various topological lower bounds [Bilge   et al., 1996, Bilge, 1995], but
they form an
overdetermined system.
In Section 2 we show that the solutions of {\it Set b} for an $N$
dimensional
gauge group, depend exactly on $N$ arbitrary constants, provided that the
system is consistent. Thus the {\it Set b} lacks the rich
structure of the self-duality equations in four dimensions.
On the other hand, the solutions of {\it Set a} do not saturate
the topological lower bounds obtained in [Bilge   et al., 1996, Bilge, 1995], but these equations form an
elliptic system under the Coulomb gauge condition.

In Section 3, we give an eigenvalue criterion to define a subvariety
 ${\cal P}_{8}$ of $8\times 8$ skew-symmetric matrices and we show that it
contains the solutions of  {\it Set a}
as a maximal linear submanifold. We
 give an eight order action whose extrema are achived on ${\cal P}_8$.

\vskip 1cm
\noindent
{\bf 2. The self-duality equations of Corrigan   et al.}
\vskip .2cm

We will study the self-duality equations (3.39) and (3.40) in [Corrigan   et al., 1983)],
that describe a scalar field $F$ which is an  eigenvector of a fourth rank
tensor $T$
invariant under $SO(7)$. We present below the two sets of equations
corresponding to the eigenvalues
$1$ and $-3$ of $T$. The first set corresponding to the eigenvalue 1 is
given below. In the following $\omega $ will denote a 2-form, and $\omega_{ij}$
will be its components with respect to an orthonormal basis.

 \noindent
{\it Set a:}
$$\eqalignno{
&\omega_{12}+\omega_{34}+\omega_{56}+\omega_{78}=0,\cr
&\omega_{13}-\omega_{24}+\omega_{57}-\omega_{68}=0,\cr
&\omega_{14}+\omega_{23}-\omega_{67}-\omega_{58}=0,\cr
&\omega_{15}-\omega_{26}-\omega_{37}+\omega_{48}=0,\cr
&\omega_{16}+\omega_{25}+\omega_{38}+\omega_{47}=0,\cr
&\omega_{17}-\omega_{28}+\omega_{35}-\omega_{46}=0,\cr
&\omega_{18}+\omega_{27}-\omega_{36}-\omega_{45}=0.&(2.1)\cr}$$
The    second set is obtained by equating the terms in each row, i.e.,

\noindent
{\it Set b:}
$$\eqalignno{
&\omega_{12}=\omega_{34}=\omega_{56}=\omega_{78},\quad ....&(2.2)\cr}$$
%&\omega_{13}=-\omega_{24}=\omega_{57}=-\omega_{68},\cr
%&\omega_{14}=\omega_{23}=-\omega_{67}=-\omega_{58},\cr
%&\omega_{15}=-\omega_{26}=-\omega_{37}=\omega_{48},\cr
%&\omega_{16}=\omega_{25}=\omega_{38}=\omega_{47},\cr
%&\omega_{17}=-\omega_{28}=\omega_{35}=-\omega_{46},\cr
%&\omega_{18}=\omega_{27}=-\omega_{36}=-\omega_{45},&(2.2)\cr}$$
%
Note that
the {\it Set a} is the orthogonal complement of {\it Set b}
with respect to the standart inner product on matrices, $\langle
A,B\rangle={\it tr}\ AB^t$.

\vskip .4cm
\noindent
{\bf 2.1. The number of free parameters in the solution of {\it Set
b}.}

\vskip .2cm

Let $F$ be the  curvature  2-form,
$F=\sum_{a,b}F_{ab}E_{ab}$ where the $E_{ab}$'s are basis vectors for the
Lie algebra of the gauge group. Assume that each 2-form $F_{ab}$ satisfies the
equations in {\it Set b}, or more generally belongs to any linear submanifold
of ${\cal S}_8\cup \{0\}$. As these equations are overdetermined, there may not
be any solutions. We recall that a topologically nontrivial solution (i.e.
where $F$ is not an exact form) is given by Grossman et. al.
[Grossman  et al., 1984]. Here we will show that, for an $N$ dimensional gauge group,  if the  field equations are consistent,
then the solution depends at most on  $N$ arbitrary constants.

The {\it Set b} represents 21 equations for the 8
components of
the connection 1-form. In addition, if we  impose the Coulomb gauge condition,
for each $F_{ab}$ we have a system of    22 equations
for 8 unknowns.
However the integrability conditions of the
  equations for the connection 1-form become quickly very cumbersome.
Thus,
instead of looking at the compatibility of the differential equations for the
connection, we look at the Bianchi identities, which are viewed as
first order differential equation for the curvature, i.e.
$$dF_{ab}=A_{ac}F_{cb}-F_{ac}A_{cb}.\eqno(2.3)$$
If each 2-form $F_{ab}$ satisfies the equations in {\it Set b} or more
generally belongs to a linear submanifold of ${\cal S}_8\cup \{0\}$, it  can be
written
as $F_{ab}=\sum_{i=1}^7F^i_{ab}h'_i$, with respect to some basis   $\{h'_i\}$
  (one set is  actually given by Eq.(2.9)). Then
$$dF_{ab}=\sum_{i=1}^7\sum_{j=1}^8\partial_jF^i_{ab}dx^jh'_i.\eqno(2.4)$$
Thus the Bianchi identities, which are 3-form equations consist of sets of 56
algebraic equations for the 56 partial derivatives $\partial _jF^i_{ab}$, for
each pair of indices $(ab)$.
 It is checked that this system is
nondegenerate, therefore
if the gauge group
is abelian, then the Bianchi identities reduce to  homogeneous equations
and    the $F_{ab}$'s  are constants. In the nonabelian case,
the Bianchi identities form an inhomogeneous system, from which all partial
derivatives of the $F_{ab}$'s are determined. Therefore, if the gauge field
equations for the connection are
consistent,  then the resulting curvature 2-forms $F_{ab}$'s can depend at
most on one arbitarary constant for each pair of indices $(ab)$. Thus we
have

\proclaim Proposition 2.1. Let $F=dA-A\wedge A$ where $A$  belongs to   an $N$
dimensional Lie algebra,  and  $F_{ab}=$ satisfy the equations in {\it
Set b}.   Then,
if the system is compatible,  $F$ can depend
at most on $N$ arbitrary constants.

\vskip .5cm
\noindent
{\bf 2.2. Ellipticity properties of {\it Set a} and {\it Set b}.}

Recall that  $F=dA-A\wedge A$, and the characteristic determinant [John, 1982] of the
 field equations are obtained using the linear part of this equation, i.e $F\sim dA$.
The Coulomb gauge condition is
$$\sum_i^{n} \partial _iA^i=0.\eqno(2.5)$$
The   characteristic determinant of the {\it Set a} together with the
Coulomb gauge condition is obtained and we have,

\proclaim Proposition 2.2.
The characteristic determinant of the {\it Set a}, together with the Coulomb
gauge condition is
$$(\xi_1^2+\xi_2^2+\xi_3^2+\xi_4^2+\xi_5^2+\xi_6^2+\xi_7^2+\xi_8^2)^4\eqno(2.6
) $$ hence the system is elliptic.

A system of elliptic equations should have as many
equations as unknowns.
The requirement of ellipticity is the injectivity and the surjectivity of the
symbol. If the symbol is injective (but not surjective), then the
system is called {\it overdetermined elliptic}.  The injectivity of the
symbol leads to certain inequalities in terms of various Sobolev norms
[Donaldson and Kronheimer, 1990]. On the other
hand the surjectivity of the symbol guarantees the solvability of the
system. Thus if a system is overdetermined elliptic, one can still use
standart results from  elliptic theory, provided that the existence of
solutions to the overdetermined system are guaranteed.

The {\it  Set b}  together with the Coulomb gauge condition is an
overdetermined
system. The subsystem consisting of the
equations
 $$\eqalign{
\omega_{12}&= \omega_{34},\quad\quad
\omega_{13} =-\omega_{24},\quad\quad
\omega_{14} = \omega_{23},\quad\quad
\omega_{15} =-\omega_{26},\cr
\omega_{16}&= \omega_{25},\quad\quad
\omega_{17} =-\omega_{28},\quad\quad
\omega_{18} = \omega_{27}\cr}\eqno(2.7)$$
together with the Coulomb gauge condition form an elliptic system.

\proclaim Proposition 2.3.
The characteristic determinant of the Eqs.(2.7) together with the
Coulomb gauge condition is
$$(\xi_1^2+\xi_2^2)
(\xi_1^2+\xi_2^2+\xi_3^2+\xi_4^2)
(\xi_1^2+\xi_2^2+\xi_3^2+\xi_4^2+\xi_5^2+\xi_6^2+\xi_7^2+\xi_8^2)
\eqno(2.8)$$
hence the {\it Set b} is overdetermined elliptic.

We note that the rank of the characteristic system of the {\it Set b}
consisting of 21 equations is 7, hence the Coulomb gauge condition is needed in
order to obtain a full rank subsystem.

\vskip .2cm
\noindent
{\bf Remark 2.4.} The characteristic matrix $A$ of the {\it Set a} satisfies
the equation $AA^t=kI$ where $t$ denotes the transpose, $I$ is the identity
matrix and
$k=
\sum_{i=1}^8\xi_i^2$.
The first row of the characteristic determinant, arising from the Coulomb gauge
condition, is the
radial vector, hence the remaining seven rows represent tangent vector fields
to $S^7$. Since $S^3$ and $S^7$ are the only parallelizable spheres,   the
equations in {\it Set a} are unique analogues of the self-duality  equations in
four dimensions, as already noted in [Corrigan   et al., 1983)].

\vskip .4cm

\noindent
{\bf 2.3 An alternative derivation of the {\it Set a} and {\it Set b}.}

We recall that squares of strongly-self dual
2-forms are self-dual in the Hodge sense [Bilge   et al., 1996)] and
the maximal linear subspaces of
strongly self-dual 2-forms are a six parameter family of 7 dimensional spaces.
In this section we will obtain analogues of
Eqs.(2.2) that will be used in Section 3. Similar equations are also obtained
in [Bilge  et al., 1995].

We fix a
nondegenerate 2-form $h'_1=e_{12}+\alpha e_{34}+\beta e_{56}+\gamma e_{78}$,
and we
consider the 2-forms $h'_j=e_{1(j+1)}+\kappa'_j$, for  $j=2,\dots 7$, such
that the
$\kappa'_j$'s do not involve $e_1$ and $e_{j+1}$. The requirement that
$(h'_1+h'_j)^2$ be
self-dual gives linear equations for the components of the $h'_j$'s. Once these
equations are solved, the non-linear equations obtained from the
self-duality of
$(h'_i+h'_j)^2$ for $i\ne 1$ can be solved easily and we obtain
the following result.

\proclaim Proposition 2.6.
Let  $h'_1=e_{12}+\alpha e_{34}+\beta e_{56}+\gamma  e_{78}$, and $h'_j$,
$j=2,\dots 7$ be of the form  $h'_j=e_{1(j+1)}+\kappa'_j$, such that
$\langle e_1,\kappa'_j\rangle=\langle e_{j+1},\kappa'_j\rangle=0$. If the
4-forms $(h'_i+h'_j)^2$ are self dual for all $i,j$ then the $h'_i$'s are
$$\eqalignno{
h'_1 &= e_{12} +\beta\gamma  e_{34} + \beta e_{56} + \gamma e_{78}  \cr
h'_2&= e_{13} - \beta\gamma e_{24} + \beta c'\ e_{57} - \beta c\  e_{58}
      - \gamma c\  e_{67} - \gamma c'\  e_{68}\cr
h'_3 &= e_{14} + \beta \gamma e_{23} - c\  e_{57} - c'\  e_{58}
      - \beta \gamma c'\  e_{67}+ \beta \gamma c\  e_{68}\cr
h'_4 &= e_{15} -\beta e_{26} - \beta c'\  e_{37}
         + \beta c\  e_{38}+ c\  e_{47}+ c'\  e_{48}\cr
h'_5 &= e_{16} + \beta e_{25} + \gamma c\  e_{37} + \gamma c'\  e_{38}
        +\beta \gamma c'\  e_{47} - \beta \gamma c\  e_{48}\cr
h'_6 &= e_{17} - \gamma e_{28} + \beta c'\  e_{35} -\gamma c\  e_{36}
       - c\  e_{45} - \beta \gamma c'\  e_{46}\cr
h'_7 &= e_{18} + \gamma e_{27} - \beta c\ e_{35} - \gamma c'\  e_{36}
       - c'\  e_{45} + \beta \gamma c\  e_{46}&(2.9)\cr} $$
where $\beta^2=\gamma^2=c^2+c'^2=1$.

Thus depending on the possible choices for
$\beta $ and $\gamma$ we have four sets of seven equations parametrized by $c$
and $c'$. We denote these forms by $h'_i$, $k'_i$,
$m'_i$ and $n'_i$ corresponding respectively to the cases
$(\beta=1,\gamma=1)$, $(\beta=1,\gamma=-1)$, $(\beta=-1,\gamma=1)$ and
$(\beta=-1,\gamma=-1)$.

  The set consisting  of the 28 forms thus obtained is however linearly
dependent for any $c$ and $c'$.
To retain similarity with Eq.(2.2) we set $c'=1$ and $c=0$, and we obtain the
following linear submanifolds of ${\cal
S}_8\cup\{0\}$.
$$\eqalignno{ B^{++}=&{\rm span}\{h'_1,h'_2,h'_3,h'_4,h'_5,h'_6,h'_7\}\cr
B^{+-}=&{\rm span}\{k'_1,k'_2,k'_3,h'_4,k'_5,k'_6,k'_7\},\cr
B^{-+}=&{\rm span}\{m'_1,m'_2,k'_3,m'_4,m'_5,m'_6,h'_7\},\cr
B^{--}=&{\rm span}\{n'_1,n'_2,h'_3,m'_4,n'_5,n'_6,k'_7\}.&(2.10)\cr}$$
A basis for 2-forms on $R^8$ can be obtained by adding
$$\eqalignno{
p'_1=& e_{14}-e_{23}+e_{58}-e_{67},\cr
p'_2=& e_{14}+e_{23}+e_{58}+e_{67},\cr
p'_3=& e_{15}+e_{26}-e_{37}-e_{48},\cr
p'_4=& e_{15}-e_{26}+e_{37}-e_{48},\cr
p'_5=& e_{18}+e_{27}+e_{36}+e_{45},\cr
p'_6=& e_{18}-e_{27}-e_{36}+e_{45},&(2.11)\cr}$$
to the 2-forms in (2.10).

The analogues of the equations in {\it Set b} are obtained by resticting
$\omega$ to the subspaces in (2.10). Similarly the analogues of {\it Set a}
are
obtained by taking orthogonal complements. The coefficients of $\omega$ with
respect to the basis consisting of the $h'_i$'s, $k'_i$'s, $m'_i$'s, $n'_i$'s
and $p'_i$'s will be denoted by the same symbols without prime.

\vskip 1cm

\noindent
{\bf 3. An eigenvalue  characterization of the {\it Set a} and an
action density.}

\vskip .2cm

We recall the following definition given in [Bilge   et al., 1996)].
\vskip .2cm
\noindent
{\bf Definition 3.1.}
Let
$\omega$ be a 2-form    in $2n$ dimensions, with
components $\omega_{ij}$ with respect to an orthonormal basis. The 2-form
$\omega$  is
called {\it strongly self-dual} if the the absolute values of the eigenvalues
of the matrix $\omega_{ij}$ are equal.
The non-zero strongly self-dual
2-forms belong to a 13 dimensional submanifold ${\cal S}_8$, and
the solutions of the {\it Set b} are among the maximal
linear submanifolds of ${\cal S}_8\cup \{0\}$
[Bilge  et al., 1995].
\vskip .2cm

We will define below a subvariety  ${\cal P}_8$ which contains
the solutions of {\it Set a} as a maximal linear
submanifold.

Let the eigenvalues of the matrix $\omega_{ij}$ be $\pm i\lambda_k$,
$k=1,\dots,4$, and define
$q_j$ to be the $j$'th elementary symmetric function of the $\lambda_k^2$'s.
Then
$$\eqalignno{
(\omega,\omega)=&s_2=4q_1=\lambda_1^2+\lambda_2^2+\lambda_3^2
                                    +\lambda_4^2,\cr
\t{1\over 2^2}(\omega^2,\omega^2)=&
s_4=6q_2=\lambda_1^2\lambda_2^2+\lambda_1^2\lambda_3^2
                   +\lambda_1^2\lambda_4^2+\lambda_2^2\lambda_3^2
                   +\lambda_2^2\lambda_4^2+\lambda_3^2\lambda_4^2,\cr
\t{1\over 6^2}
(\omega^3,\omega^3)=&s_6=4q_3=\lambda_1^2\lambda_2^2\lambda_3^2
                +\lambda_1^2\lambda_2^2\lambda_4^2
                +\lambda_1^2\lambda_3^2\lambda_4^2
                +\lambda_2^2\lambda_3^2\lambda_4^2, \cr
\t{1\over 24^2}(\omega^4,\omega^4)=&s_8=q_4=
               \lambda_1^2\lambda_2^2\lambda_3^2\lambda_4^2.
            &(3.1) \cr}$$
We have the inequalities
$$q_1^2\ge q_2\ge \sqrt{q_4},\eqno(3.2)$$
 the equalities
being saturated iff all the eigenvalues are equal [10], i.e. for the strongly
self-dual forms. This corresponds to the case where the quantities
$$\eqalign{A&=(\omega,\omega)^2-\t{2\over 3}(\omega^2,\omega^2),\cr
           B&=(\omega^2,\omega^2)-(\omega^4,\omega^4)^{1/2}\cr}\eqno(3.3)$$
vanish. The proposition 3.2 below implies that the quantity
$$\Phi=A+{1\over 3} B=(\omega  ,\omega  )^2-{1\over
3}(\omega^2,\omega^2)-{1\over 3}(\omega^4,\omega^4)^{1/2}\eqno(3.4)$$
is a measure of the power of the anti self-dual part of $\omega$.

\proclaim Proposition 3.2. Let  $(\omega^{2+},\omega^{2+})\ge
(\omega^{2-},\omega^{2-})$, and
$\Phi=(\omega,\omega)^2-\alpha(\omega^{2+},\omega^{2+})$, where
$\omega^{2\pm}$
denote the self-dual and anti self-dual parts of $\omega^2$. Then, ${\rm
max}\ \alpha$
such that $\Phi$ is non-negative for all $\omega$ is $\alpha={3\over 2}$.

\noindent
{\it Proof.} If
$(\omega^{2+},\omega^{2+})\ge
(\omega^{2-},\omega^{2-})$, then  $(\omega^4,\omega^4)^{1/2}=*\omega^4=
(\omega^{2+},\omega^{2+})- (\omega^{2-},\omega^{2-})$.
From the inequalities (3.2), it can be seen that $\alpha\le
3/2$.
On the other hand the equality is attaigned for $\omega\in {\cal S}_8$, hence
$\alpha=3/2$. \quad e.o.p.

\vskip .2cm

It is an elementary fact that the product ${1\over 3} AB$, under the constraint
$A+{1\over 3}B=$const. is maximized for $\Psi=A-{1\over 3}B=0$, and minimized
for $A=0$ or $B=0$. The condition $A-{1\over 3}B=0$  gives
$$\Psi=(\omega,\omega)^2-(\omega^2,\omega^2)
+{1\over 3}(\omega^4,\omega^4)^{1/2}=0.\eqno(3.5)$$
Thus we have

\proclaim Proposition 3.3. Let
$\Phi=(\omega,\omega  )^2-{1\over
3}(\omega^2,\omega^2)-{1\over 3}(\omega^4,\omega^4)^{1/2}$ be fixed and
$(\omega,\omega)^2-{2\over 3} (\omega^2,\omega^2)$ be nonzero. Then the
quantity
$(\omega,\omega)^2-\t{2\over 3}(\omega^2,\omega^2)
(\omega^2,\omega^2)-(\omega^4,\omega^4)^{1/2}$ is maximal for
$\Psi=(\omega,\omega)^2-(\omega^2,\omega^2)
+{1\over 3}(\omega^4,\omega^4)^{1/2}=0$.

The expression of $\Psi$ in terms of the $\omega_{ij}$'s is very complicated.
However it reduces to a relatively simple form under a change of parameters.
If we reparametrize the eigenvalues as
$$\eqalignno{
\epsilon_1=&(\lambda_1+\lambda_2+\lambda_3+\lambda_4),\cr
\epsilon_2=&(\lambda_1-\lambda_2-\lambda_3+\lambda_4),\cr
\epsilon_3=&(\lambda_1-\lambda_2+\lambda_3-\lambda_4),\cr
\epsilon_4=&(\lambda_1+\lambda_2-\lambda_3-\lambda_4),&(3.6)\cr}$$
then $\Psi$ reduces to
$$\Psi=(\omega,\omega)^2-(\omega^2,\omega^2)  +{1\over
3}(\omega^4,\omega^4)^{1/2}=\epsilon_1\epsilon_2\epsilon_3\epsilon_4.\eqno(3.7)
$$
We note that we could obtain a similar decomposition for
$(\omega,\omega)^2-(\omega^2,\omega^2)  -{1\over
3}(\omega^4,\omega^4)^{1/2}$,
if we defined the $e_i$'s with an odd number of minus signs.

The equality of the $\lambda_i$'s corresponds to the case where any three of
the $\epsilon_i$'s are zero. The appropriate nonlinear set containing solutions
of {\it Set a} is thus the set where only one of the $\epsilon_i$'s is zero.
The explict expression of $\Psi$
when $\omega$ is written with respect to the basis given in
(2.10) and (2.11) is given below.
$$\eqalignno{\Psi=
h_1 [ &k_1 (m_1 n_1 + m_4 p_3 + m_5 n_5)
     + k_2 (n_1 m_2 - m_4 p_6 + m_5 n_6)
     + k_3 (n_1 p_1 - p_3 n_6 - n_5 p_6)\cr
     +&k_6 (m_1 n_6 + m_4 p_1 - n_5 m_2)
     + k_7 (m_1 p_6 + p_3 m_2 + m_5 p_1)\cr
+h_2 [& k_1 (m_1 n_2 + m_4 p_5 + n_5 m_6)
     + k_2 (m_4 p_4 + m_2 n_2 + n_6 m_6)
     + k_3 (n_5 p_4 - n_6 p_5 + p_1 n_2)\cr
     +&k_5 ( - m_1 n_6 - m_4 p_1 + n_5 m_2)
     + k_7 ( - m_1 p_4 + m_2 p_5 + p_1 m_6)\cr
+h_3 [& k_1 (m_1 p_2 + p_3 m_6 - m_5 p_5)
     + k_2 ( - m_5 p_4 + m_2 p_2- p_6 m_6)
     + k_3 (p_3 p_4 + p_6 p_5 + p_1 p_2) \cr
     +&k_5 (m_1 p_6 + p_3 m_2 + m_5 p_1)
     + k_6 ( - m_1 p_4 + m_2 p_5 + p_1 m_6)\cr
+h_4 [& m_1 (n_1 p_4 + p_6 n_2 + n_6 p_2)
     + m_2 ( - n_1 p_5 + p_3 n_2- n_5 p_2)
     + m_4 (p_3 p_4 + p_6 p_5 + p_1 p_2)\cr
     +&m_5 (n_5 p_4 -n_6 p_5 + p_1 n_2)
     + m_6 ( - n_1 p_1 + p_3 n_6 + n_5 p_6)\cr
+h_5[ & k_2 ( - n_1 m_6 + m_4 p_2 + m_5 n_2)
     + k_3 (n_1 p_5 - p_3 n_2+ n_5 p_2)
     + k_5 (m_1 n_1 + m_4 p_3 + m_5 n_5) \cr
     +&k_6 (m_1 n_2 +m_4 p_5 + n_5 m_6)
     + k_7 ( - m_1 p_2 - p_3 m_6 + m_5 p_5)\cr
+h_6 [& k_1 (n_1 m_6 - m_4 p_2 - m_5 n_2)
     + k_3 (n_1 p_4 + p_6 n_2 +n_6 p_2)
     + k_5 (n_1 m_2 - m_4 p_6 + m_5 n_6) \cr
     +&k_6 (m_4 p_4 + m_2n_2 + n_6 m_6)
     + k_7 (m_5 p_4 - m_2 p_2 + p_6 m_6)\cr
+h_7 [& k_1 (n_1 p_5 - p_3 n_2 + n_5 p_2)
     + k_2 (n_1 p_4 + p_6 n_2 +n_6 p_2)
     + k_5 ( - n_1 p_1 + p_3 n_6 + n_5 p_6)\cr
     +&k_6 ( - n_5 p_4+ n_6 p_5 - p_1 n_2)
     + k_7 (p_3 p_4 + p_6 p_5 + p_1 p_2)&(3.8)\cr}$$

From (3.8), it can be seen that $\Psi=0$ both on  the {\it Set a} where all
all $h_i$'s are zero, and on the {\it Set b} where all the parameters except
the $h_i$'s are zero. Actually
$\Psi$ vanishes on
the complement of each of the subspaces in (2.10), which are
21 dimensional linear submanifolds of ${\cal P}_8$.
 By assigning
arbitrary values to the remaining parameters, it can be seen that these 21
dimensional
submanifolds are maximal. Hence solutions of {\it Set a} (and their analogues)
are among the maximal linear submanifolds of ${\cal P}_8$.

\vskip 1cm
\noindent
{\bf Acknowledgements.} This work is partially supported by the Turkish
Scientific and Technological Research Council, TUBITAK.
\vskip 1cm

\vskip 1cm
\baselineskip 14pt

\noindent
{\bf References.}
\vskip .3cm

\item{} A.H. Bilge, T. Dereli and S. Kocak, ``An explicit
construction of self-dual 2-forms in eight dimensions", hep-th
9509041 (1995).
\vskip .2cm

\item{} A.H. Bilge, T. Dereli and S. Kocak, ``Self-dual Yang-Mills fields
in eight dimensions", {\it Lett. Math. Phys.}, {\bf 36}, 301, (1996).
\vskip .2cm

\item{} A.H. Bilge, ``Self-duality in demensions $2n>4$: equivalence of various
definitions and an upper bound for $p_2$. preprint: dg-ga 9604002.
\vskip .2cm

\item{}E.Corrigan, C.Devchand, D.B.Fairlie and J.Nuyts, ``First-order
equations for gauge fields in spaces of dimension greater than four", {\it
Nuclear Physics} {\bf B214}, 452-464, (1983).
\vskip .2cm

\item {} S.K. Donaldson and P.B. Kronheimer, {\bf The Geometry of Four
Manifolds},  Clarendon, Oxford (1990).
\vskip .2cm

\item{}B.Grossman, T.W.Kephart, J.D.Stasheff, ``Solutions to Yang-Mills
field equations in eight dimensions and the last Hopf map", {\it Commun. Math.
Phys.}, {\bf 96}, 431-437, (1984).
\vskip .2cm

\item{} F. John, {\bf Partial Differential Equations}, Springer-Verlag, New
York, (1982).
\vskip .2cm

\item{} M.Marcus anf H. Minc, {\bf A Survey of Matrix Theory and Matrix
Inequalities}, Dover Publ. New York, (1964).
\vskip .2cm

\item{} A. Trautman, {\it Inter. J. Theor. Phys.}, {\bf 16}, 561-656,
(1977).
\vskip .2cm

\end